# Growth of whisker-like and bulk single crystals of PrFeAs(O,F) under high pressure

N. D. Zhigadlo

*Laboratory for Solid State Physics, ETH Zurich, CH-8093 Zurich, Switzerland*

**Abstract**

Superconducting whisker-like and bulk single crystals of PrFeAs(O,F) were grown at 3 GPa and ~ 1500 ºC from NaAs flux using the cubic anvil high-pressure and high-temperature technique. The ribbon-shaped whisker-like crystals were 400 – 1300 μm in length, 40 – 70 μm in width, and 5 – 10 μm in thickness with the long sharp edge in the *ab*-plane and the *c*-axis perpendicular to the face. The grown bulk crystals were found to have irregular plate-like shapes with linear sizes up to 1 mm and 120 μm in thickness. The both type of single-crystalline objects are crystallized in the same tetragonal structure and showed a single-phase behavior in the temperature dependence of magnetization with a superconducting transition temperature of about 31 K. The presence of temperature gradient and reactive flux seems to play an important role for the growth of single-crystalline whiskers and bulk single crystals of PrFeAs(O,F). These observations demonstrate that the morphological properties of the crystals are depending not only on the symmetry of the crystal structure of the desired material, but also of the growth medium.

---

*E-mail address:* zhigadlo@phys.ethz.ch (N. D. Zhigadlo)

# 1. Introduction

While the phenomenon of superconductivity in the Fe-based pnictide materials continues to be an open question, studies on good quality single crystals are of fundamental importance to understand the basic mechanism and for the elucidation of the layered crystal structure in relation to the intrinsic and anisotropic superconducting properties [see Ref. 1 and references therein]. Since the discovery of the Fe-based superconductors, much effort has been focused on the growth of single-crystalline samples. Among the all known Fe-based families, the quaternary iron oxypnictides ($Ln$1111 – $Ln$Fe$Pn$O, $Ln$: lanthanide, $Pn$: pnictogen) look very interesting because it still hold the highest superconducting critical temperature ($T_c$) of ~ 55 K and has several unique and promising for practical application properties [1-3]. In contrast to other families, the growth crystals of multicomponent $Ln$1111 system is far more complex, resulting in a higher degree of synthesis difficulty and a lower level of growth control [4-8]. Despite extensive experimental efforts over the past years, the $Ln$1111 crystals grown by various methods are still limited in dimensions and the growth mechanism is not yet satisfactory understood. Since the $Ln$1111 compounds melts incongruently the application of flux method was found suitable for growing single crystals. In flux method the components of the desired substance are dissolved in a solvent and the growth is preceded at the temperature well below the melting point of the solute phase. In the case of flux growth at ambient pressure in quartz ampoules the growth temperature is limited and thus the solubility of starting components is very low [9]. We have adopted the high-pressure and high-temperature (HPHT) method for growth of $Ln$1111 crystals using NaCl-KCl as a flux [4-7]. This HPHT flux method avoids vaporization loses of volatile elements, increases the solubility of the components, and improves to a certain degree control of composition without being perfect. After systematic investigations of all parameters influencing on the crystal growth, we determined that the growth temperature and soaking time are most important parameters influencing the crystal size. However, while this method allows growing $Ln$1111 crystals with high $T_c$ (≥50 K), even at optimal HPHT conditions, the size of grown crystals was limited by ~ 300 μm in length and ~ 20 μm in thickness mainly due to the low growth rate and high density of nucleation sites [4-7]. To make crystals of larger sizes, further improvements of the growth conditions are required. Yan et al. [10] reported that NaAs flux has some advantages for increasing the size of *Ln*1111 crystals, but the dopant content has yet to be controlled. Recently, we also found that in comparison with NaCl-KCl flux-method, the high-pressure and high-temperature NaAs (KAs) flux-growth technique are at least three times more efficient in obtaining large size crystals [8]. However, because NaAs (KAs) now acting not only like a flux but also as a component of the system providing extra As, it is become more difficult to control the *Ln*1111 crystal growth process. During our efforts on growing bulk crystals of the Pr1111 system we observed formation of crystals with peculiar ribbon-shaped geometry. Such unusual morphology is not very typical for these pnictide superconductors and very likely related to the specific conditions of the HPHT growth. Among Fe-based superconductors the first one-dimensional F-doped SmFeAsO nanocables were fabricated using ZnO nanotubes as a template [11]. Recently, superconducting $Ca_{10}(Pt_4As_8)(Fe_{1.8}Pt_{0.2}As)_5$ nanowhiskers with widths down to hundreds of nanometers were grown in a Ta capsule in an evacuated quartz tube by a flux method [12].

In this paper, we report details of successful HPHT growth of superconducting PrFeAs(O,F) crystals. We demonstrate that the application of 3 GPa pressure and 1500 °C growth temperature allows growing not only bulk crystals but also elongated ribbon-shaped whisker-like crystals. An appearance of such kind of whiskers reveals that the morphological properties of single-crystalline objects are dependent not only on the symmetry of the crystal structure of desired material, but also on the environmental conditions in which the crystals are grown.

## 2. Experimental details

For the growth of PrFeAs(O,F) crystals, we used the cubic anvil HPHT technique, which was already developed earlier in our laboratory for growing superconducting crystals and various other compounds [13-16]. The nominal composition of the precursor was $PrFeAsO_{0.7}F_{0.3}$. The starting materials of PrAs, $FeF_2$ (or $PrF_3$), FeO (or $Fe_2O_3$), and Fe of high purity ($\geq$ 99.95 %) were weighed according to the stoichiometric ratio, thoroughly mixed using a mortar, and mixed with NaAs. Arsenide of NaAs was prepared by reacting Na metal with As pieces in a boron nitride (BN) container at pressure of 3 GPa and temperature of 600 °C for 4

h. The precursor to NaAs flux molar ratio was varied between 1:1 and 1:3. After optimization, we found the mixture of Pr1111 precursor with NaAs flux in the molar ratio 1:1 to be most effective for growing Pr1111 crystals. At such ratio the 1111-type phase is still main dominant product, whereas too-high precursor-to-flux ratio prevents the Pr1111 phase formation and results in an increasing amount of impurity phases. We note that the synthesis temperature (~ 1500 ºC in present work) and soaking time are also very important parameters for growing sizable crystals. At such "extreme" conditions often the 1111-type phase tends to decompose into various phases. A further understanding and optimization of crystal growth process is still necessary in order to grow optimally doped $Ln$1111 crystals with high $T_c'$s. All procedures with the sample preparation were performed in a glove box due to the toxicity of arsenic and hydroscopicity of NaAs. A pellet containing precursor and flux of about 1 g in weight was enclosed in BN container of 8.0 mm in inner diameter and 10 mm in length and placed inside a pyrophyllite cube with a graphite heater.

Figure 1 shows the schematic illustration of the sample cell assembly and typical heat treatment process of the high-pressure flux growth. The sample cell assembly [Fig. 1 (a)] consisted of a cubic block of pyrophyllite as a pressure-transmitting medium, a graphite tube as a heater with electrical leads made of discs of stainless steel. The BN crucible fits inside the graphite tube and the sample of starting materials was placed inside the BN crucible. An apparatus with six anvils of tungsten carbide compressed the cubic cell to 3 GPa at room temperature. The crystal growth process was conducted by heating up a mixture of $PrFeAsO_{0.7}F_{0.3}$ solute and NaAs flux to the maximum temperature of ~ 1500 °C in 2 h, kept for 5 h, cooling to 1250 °C by applying a rate of 2.8 – 5 °C h$^{-1}$, holding at this temperature for 3 h, and then finally cooling down to room temperature in 2 h [Fig. 1 (b)]. After completing the crystal growth process the crystalline products were then separated by dissolving the flux in distilled water. As shown schematically in Fig. 1 (a) by dashed lines, the temperature gradient across the sample is an important parameter that influences the crystal growth process. Sample temperature is estimated by the predetermined relation between applied electrical power and measured temperature in the cell. The temperature calibration was conducted at the central, upper, and lower regions of the furnace. For the cell temperature of ~ 1450 ºC, we estimated the temperature gradient across the sample to be around 70 ºC. The growth temperature denoted in present study corresponds to the temperature of the central position of the furnace where the BN crucible was fixed, as schematically illustrated in Fig. 1(a).

The general morphology and dimensionality of the grown single-crystalline objects was observed by an optical microscope (Leica M 205 C). The x-ray quality of the crystals and the growth directions were studied at room temperature using a Bruker diffractometer with CuKα radiation. The elemental analysis of the grown single-crystalline products was performed by means of energy dispersive x-ray (EDX) spectrometry. Temperature-dependent magnetization measurements were carried out with a Quantum Design Magnetic Property Measurement System (MPMS) with the reciprocating sample option installed.

## 3. Results and discussion

The views of the general morphology of the high-pressure products of the top and bottom parts [marked as SC in Fig. 1 (a)] are shown in Fig. 2(a) and 2(b). Two main morphological types of crystalline objects were observed: the bulk plate-like PrFeAs(O,F) crystals and green pieces of PrOF impurity compound with irregular shape. These results are in good agreement with those reported by us in Ref. [8], where more detailed description is given. However, in the present study the synthesis temperature was slightly higher than in Ref. [8] and as results we obtained bigger quantity of larger crystals, while the $T_c$ is still low. Interestingly, that after dissolving the central part of the crucible, the ribbon-like single-crystalline whiskers was observed (marked as W in Fig. 3). Figure 3 shows an optical image of PrFeAs(O,F) whiskers with various lengths (L), widths (V), and thicknesses (T): L ~ 400 – 1300 μm, V ~ 40 – 70 μm, T ~ 5 – 10 μm. To our best knowledge, here we report, for the first time, the whisker growth of *Ln*1111 family. Similarly to definition given in Ref. [17], we used the notation "whisker" mainly based on the difference in the aspect ratio of the straight crystal objects, i.e. the ratio between the length (L) and width (V) taken in the *ab*-plane. If the ratio R = L/V >> 1, the crystal is defined as whisker, and if R ≈ 1 the shape is considered to be conventional crystal. Several tens of Pr1111 whiskers were aligned at the surface of sample holder and attached by vacuum grease. An x-ray diffraction pattern taken from the flat surface is shown in Fig. 4. The x-ray analysis confirmed that the grown whiskers belong to the 1111-type structure and they are single crystals. The presence of diffraction lines with the (00*l*) indices indicates that the flat surface is the *ab*-plane and the *c*-axis perpendicular to the face. Some tiny peaks other from the (00*l*) are also visible in Fig. 4. Since for this measurement many small whiskers were chosen, it is very likely that the surface of the whiskers was not completely clean off from attached impurity phases. The full width at half maximum (FWHM) of grown PrFeAs(O,F) whiskers was determined to be ~ 0.17 ° from the CuK$_{\alpha 1}$ diffraction peak of (006). This value of FWHM is comparable with those observed for high-quality of NdFeAs(O,F) [18] and SmFeAs(O,F) [19] single crystals grown under high pressure conditions. Nevertheless, it can be noted, that the FWHM of an x-ray peak is not a characteristic parameter of any crystal, since it is a combined effect of peak broadening due to material properties as well as instrumental broadening due to spectral width of the x-ray source. By using the (007) diffraction peak the lattice constant *c* was estimated to be 8.5843(8) Å, which is almost the same as the reported value [5]. In overall, there are no essential differences between details of the structure refinement for the PrFeAs(O,F) whisker-like and bulk single crystals and they are consistent with the results of our previous x-ray diffraction studies [see for details Table 1 in Ref. 5]. The resulting stoichiometries were investigated by EDX spectroscopy analysis. The EDX spectra were collected from several points and compositional analysis confirmed that the ratio of praseodymium, iron and arsenic is equal to 1:1:1. The light elements of oxygen and fluorine cannot be measured accurately; therefore, we could not determine the actual doping level of PrFeAs(O,F) crystals and whiskers.

Figures 5(a), (b) and (c) present the temperature dependences of the magnetic moment measured in low magnetic field applied along the *c*-axis for one bulk plate-like PrFeAs(O,F) single crystal, for ten oriented single-crystalline whiskers collected from a single growth batch, and for one individual whisker. The observed curves revealed a superconducting transition temperature of ~ 31 K indicating that the crystals are underdoped [20]. The width of superconducting transition in all cases is somewhat broadened, which is possibly due to slight variation in the fluorine content. A small ratio of field cooled to zero-field cooled magnetization is characteristic of superconductors with strong pinning.

Before providing an explanation of whiskers appearance in our experiments, we note, that the mechanism of spontaneous formation and growth of whiskers represent a controversially discussed phenomenon for more that fifty years [21]. In practice, metal and ceramic whiskers often grew from the vapor phase by a spiral growth mechanism in which a crystal face with screw dislocation can add on new material much faster than other crystal faces, resulting in extreme crystal elongation in the direction of rapid growth [21]. Later on it was recognized that in many cases spiral growth mechanism cannot be invoked as the cause of crystal elongation. It appears that whiskers can grow by so-called a Vapor-Liquid-Solid (VLS) mechanism, in which a drop of liquid at the tip of a whisker controls growth. Atoms are transferred along the crystal surface to the liquid droplet and then crystallize into the growing whisker tip [21, 22]. For more complex oxide materials, it was realized that the mechanism of growth and the resulting morphology of crystalline objects depend on the environmental conditions. Wanklyn [23] proposed the general flux (or self-flux) growth criteria, observing the relationship between the starting composition (type of compounds and their ratio), the growth temperature and the habit of many complex as-grown crystals. For example, for anisotropic oxide materials such as tetragonal ones, depending on the environmental conditions, he obtained both straight whiskers and plate-like single crystals. In the case of the high temperature superconducting single crystal whiskers, the Bi-Sr-Ca-Cu-O system is most popular, probably due to relatively easy fabrication [17, 24-27]. The whiskers were also observed in $RBa_2Cu_3O_{7+x}$ systems with R = rare earth elements [17, 28, 29], whereas little is known about whisker growth and their properties in the Tl- and Hg-based superconductors [30]. For the case of Bi2212 and Bi2223 cuprate superconductors various mechanisms were proposed, as summarized in review by Badica et al. [27]. In many reports the authors notice that the partial melting of the compounds is an important factor which governed whisker growth. Natural question arise what are the driving forces for the growth of Pr1111 single-crystalline objects with such peculiar dimensions. Before going into details, first, let us consider the *Ln*FeAsO system from the crystal chemistry viewpoint [31, 32]. The *Ln*FeAsO are composed of both insulating *Ln*O and superconducting FeAs layers that are weakly bonded and stacked in the *c*-axis direction. The crystal structure is anisotropic and the chemical bonding strength between the *a*-axis and *c*-axis differs significantly [31, 32]. This is supported by our previous [4-8] and present data showing that the shortest dimensions always corresponds to the *c*-axis of the crystals and whiskers, i.e. the easy growth direction is along the *ab*-plane. Therefore, the plate-like morphology of *Ln*1111 single-crystalline objects with the *c*-axis along the thickness and the largest face coinciding with the *ab*-plane reflects the crystal chemistry anisotropic features. Considering that the crystal structure and crystal chemistry for straight whiskers and for plate crystals is the same, means that the only difference in the growth rates in the *a*- and *c*-directions. Thinking about the possible mechanisms of the growth, here we note a few particular details. We assume that a longitudinal temperature gradient and a high crystallization rate in the *ab*-plane play a major role in the growth. As shown schematically in Fig. 1(a), for the cell temperature of ~ 1500 °C, the estimated temperature difference between the central and outer parts of the BN crucible was around 70 °C. In the crystal growth process, a temperature gradient is maintained in a crucible and results in the transfer of species through the liquid phase from the hot central part to the cold ends. The bulk plate-like Pr1111 crystals were found at these coolest parts, whereas most of flat ribbon-shaped whiskers with rectangular cross-section and relatively sharp edges were grown at the center of crucible. Thus, we believe, that the central melting zone with a high concentration of reactive NaAs flux has an influence on the whisker growth, because the growth of whiskers needs liquid for fast ionic diffusion. Up to this level of discussion we no need to introduce any specific growth mechanism and it is possible to state that the environmental conditions influence the morphology of these single-crystalline objects. Some of them are of intrinsic type and some of extrinsic, indirectly influencing the intrinsic ones.

## 4. Conclusions

Superconducting straight ribbon-shaped whiskers and bulk plate-like single crystals of PrFeAs(O,F) oxypnictide were successfully grown by high-pressure and high-temperature method using NaAs as a flux. Both morphological types of crystals are crystallized in a tetragonal 1111-type structure with a superconducting transition temperature of 31 K. The presence of temperature gradient and reactive flux play an important role in the growth. The observed results suggest that it is the growth conditions that determine what kind crystals morphology will occur, rather than the specific mechanisms of whisker growth. In our experimental environments such conditions are attained at the central part of the crucible where whiskers form. From a practical viewpoint, further tuning of the growth rate, leading to a particular morphology, will depend on the environmental conditions.

**Acknowledgments**

This work was partially supported by the National Center of Competence in Research MaNEP (Materials with Novel Electronic Properties).

**References**

[1] S. Fujitsu, S. Matsuishi, H. Hosono, Intern. Mat. Rev. 57 (2012) 311.

[2] P. J. W. Moll, R. Puzniak, F. Balakirev, K. Rogacki, J. Karpinski, N. D. Zhigadlo, B. Batlogg, Nat. Mater. 9 (2010) 628.

[3] P. J. W. Moll, L. Balicas, V. Geshkenbein, G. Blatter, J. Karpinski, N. D. Zhigadlo, B. Batlogg, Nat. Mater. 12 (2013) 134.

[4] N. D. Zhigadlo, S. Katrych, Z. Bukowski, S. Weyeneth, R. Puzniak, J. Karpinski, J. Phys.: Condens Matter 20 (2008) 342202.

[5] J. Karpinski, N. D. Zhigadlo, S. Katrych, Z. Bukowski, P. Moll, S. Weyeneth, H. Keller, R. Puzniak, M. Tortello, D. Daghero, R. Gonnelli, I. Maggio-Aprile, Y. Fasano, Ø. Fischer, K. Rogacki, B. Batlogg, Physica C 469 (2009) 370.

[6] N. D. Zhigadlo, S. Katrych, S. Weyeneth, R. Puzniak, P. J. W. Moll, Z. Bukowski, J. Karpinski, H. Keller, B. Batlogg, Phys. Rev. B 82 (2010) 064517.

[7] N. D. Zhigadlo, S. Katrych, M. Bendele, P. J. W. Moll, M. Tortello, S. Weyeneth, V. Yu. Pomjakushin, J. Kanter, R. Puzniak, Z. Bukowski, H. Keller, R. S. Gonnelli, R. Khasanov, J. Karpinski, B. Batlogg, Phys. Rev. B 84 (2011) 134526.

[8] N. D. Zhigadlo, S. Weyeneth, S. Katrych, P. J. W. Moll, K. Rogacki, S. Bosma, R. Puzniak, J. Karpinski, B. Batlogg, Phys. Rev. B 86 (2012) 214509.

[9] L. Fang, P. Cheng, Y. Jia, X. Zhu, H. Luo, G. Mu, C. Gu, H.-H. Wen, J. Cryst. Growth 311 (2009) 358.

[10] J.-Q. Yan, S. Nandi, J. L. Zaretsky, W. Tian, A. Kreyssig, B. Jensen, A. Kracher, K. W. Dennis, R. J. McQueeney, A. I. Goldman, R. W. McCallum, T. A. Lograsso, Appl. Phys. Lett. 95 (2009) 222504.

[11] S.-M. Zhou, S. Li, P. Wang, D.-Q. Zhao, H.-C. Gong, B. Zhang, M. Wang, and X.-T. Zhang, Supercond. Sci. Technol. 21 (2008) 125007.

[12] J. Li, J. Yuan, D.-M. Tang, S.-B. Zhang, M.-Y. Li, Y.-F. Guo, Y. Tsujimoto, T. Hatano, S. Arisawa, D. Golberg, H.-B. Wang, K. Yamaura, and E. Takayama-Muromachi, J. Am. Chem. Soc. 134 (2012) 134.

[13] J. Karpinski, N. D. Zhigadlo, S. Katrych, R. Puzniak, K. Rogacki, R. Gonnelli, Physica C 456 (2007) 3.

[14] N. D. Zhigadlo, J. Karpinski, Physica C 460-462 (2007) 372.

[15] N. D. Zhigadlo, J. Karpinski, S. Weyeneth, R. Khasanov, S. Katrych, P. Wägli, H. Keller, J. Phys. Conf. Ser. 97 (2008) 012121.

[16] R. T. Gordon, N. D. Zhigadlo, S. Weyeneth, S. Katrych, and R. Prozorov, Phys. Rev. B 87 (2013) 094520.

[17] P. Badica, A. Agostino, M. M. R. Khan, S. Cagliero, C. Plapcianu, L. Pastero, M. Truccato, Y. Hayasaka, and G. Jakob, Supercond. Sci. Technol. 25 (2012) 105003.

[18] P. Cheng, H. Yang, Y. Jia, L. Fang, X. Zhu, G. Mu, and H.-H. Wen, Phys. Rev. B 78 (2008) 134508.

[19] H. S. Lee, J. H. Park, J. Y. Lee, J. Y. Kim, N. H. Sung, T. Y. Koo, B. K. Cho, C. U. Jung, S. Saini, S. J. Kim, and H. J. Lee, Supercond. Sci. Technol. 22 (2009) 075023.

[20] C. R. Rotundu, D. T. Keane, B. Freelon, S. D. Wilson, A. Kim, P. N. Valdivia, E. Bourret-Courchesne, and R. J. Birgeneau, Phys. Rev. B 80 (2009) 144517.

[21] J. D. Katz (2001) Whiskers. In: Encyclopedia of Materials: Science and Technology, Elsevier Science Ltd. p. 9571.

[22] D. R. Veblen, and J. E. Post, Am. Miner. 68 (1983) 790.

[23] B. M. Wanklyn, J. Cryst. Growth. 37 (1977) 334.

[24] I. Matsubara, H. Tanigawa, T. Ogura, H. Yamashita, and M. Kinoshita, Phys. Rev. B 45 (1992) 7414.

[25] R. Jayavel, C. Sekar, P. Murugakoothan, C. R. Venkateswara Rao, C. Subramanian, and P. Ramasamy, J. Cryst. Growth. 131 (1993) 105.

[26] M. Nagao, M. Sato, H. Maeda, S.-J. Kim, and T. Yamashita, Appl. Phys. Lett. 79 (2001) 2612.

[27] P. Badica, K. Togano, S. Awaji, K. Watanabe, H. Kumakura, Supercond. Sci. Technol. 19 (2006) R81.

[28] C. Klemenz, and H. J. Scheel, J. Cryst. Growth. 203 (1999) 534.

[29] A. T. M. N. Islam, Y. Tachiki, S. Watauchi, and I. Tanaka, Supercond. Sci. Technol. 18 (2005) 1238.

[30] A. Kikuchi, K. Inoue, and K. Tachikawa, Physica C 337 (2000) 180.

[31] W. Jeitschko, B. I. Zimmer, R. Glaum, L. Boonk, U. C. Rodewald, Z. Naturforsch. 63b (2008) 934.

[32] R. Pöttgen, D. Johrendt, Z. Naturforsch. 63b (2008) 1135.

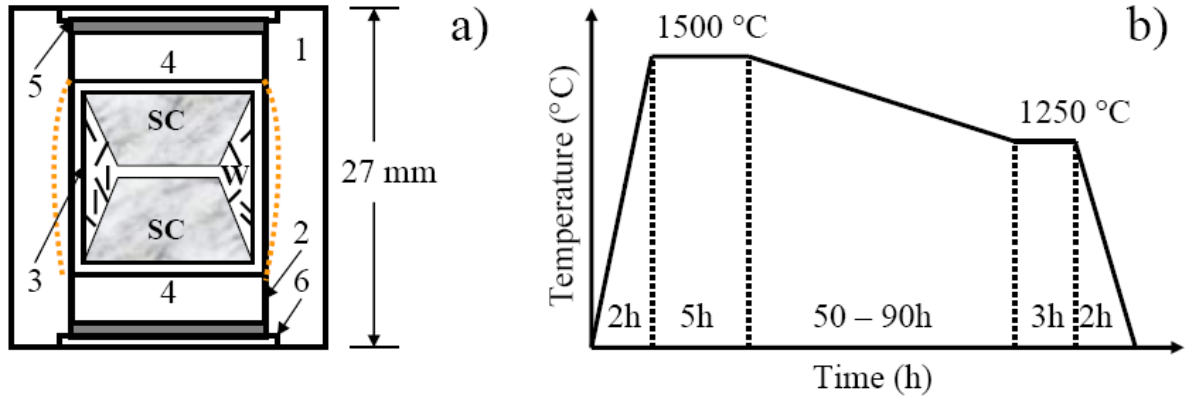

**Fig. 1.** a) Schematic illustration of the sample cell assembly for the high-pressure crystal growth: 1) Pyrophyllite cube, 2) heater-graphite sleeve, 3) BN sample crucible, 4) separator-pyrophyllite pellets, 5) electrode-graphite disks, 6) electrode-stainless steel disks. Dashed lines show the temperature gradient in high-pressure cell assembly. Pieces with collection of plate-like single crystals (SC) were found at the top and bottom parts of the crucible after dissolving of NaAs flux in water. The single-crystalline whiskers (W) were found at the center part of the crucible. b) Typical heat treatment process of the high-pressure flux growth.

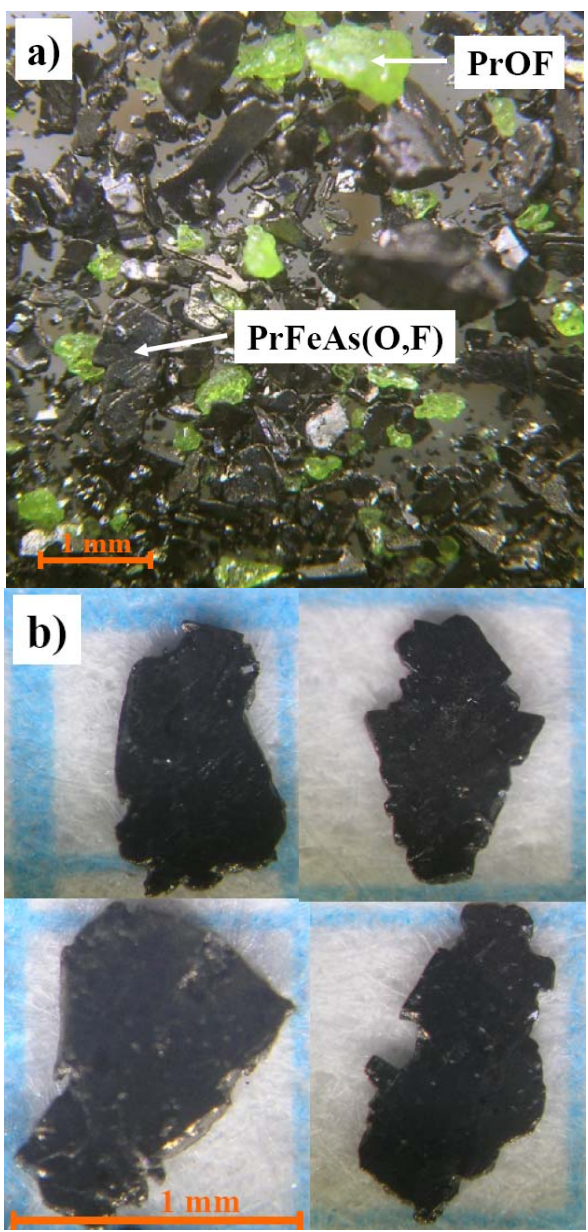

**Fig. 2.** a) View of the general morphology of high-pressure crystalline products after dissolving the flux in water: bulk plate-like PrFeAs(O,F) crystals and PrOF impurity compound are two dominant objects. b) Individual PrFeAs(O,F) crystals are shown on 1-mm grid.

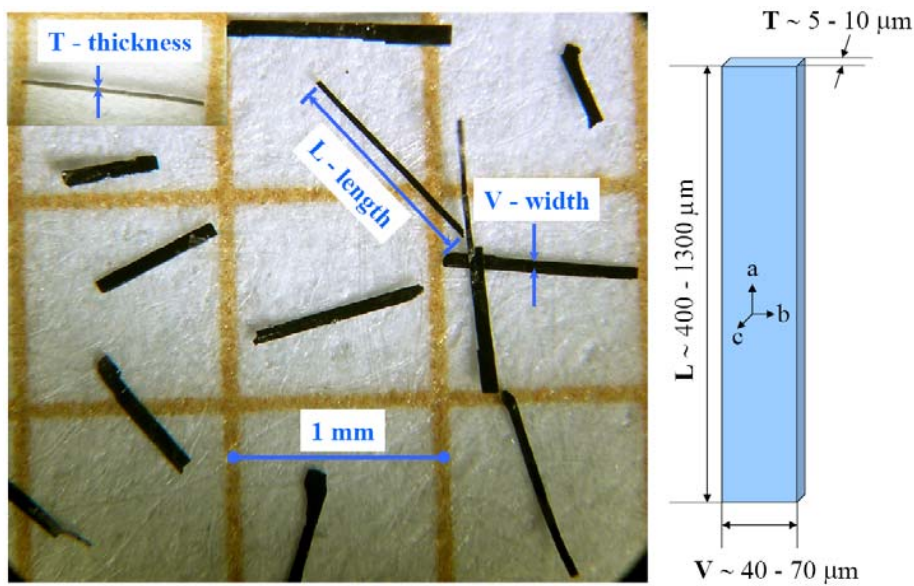

**Fig. 3.** Optical image of PrFeAs(O,F) single-crystalline whiskers. On the right side the orientation of the crystal axis and typical dimensionalities is given.

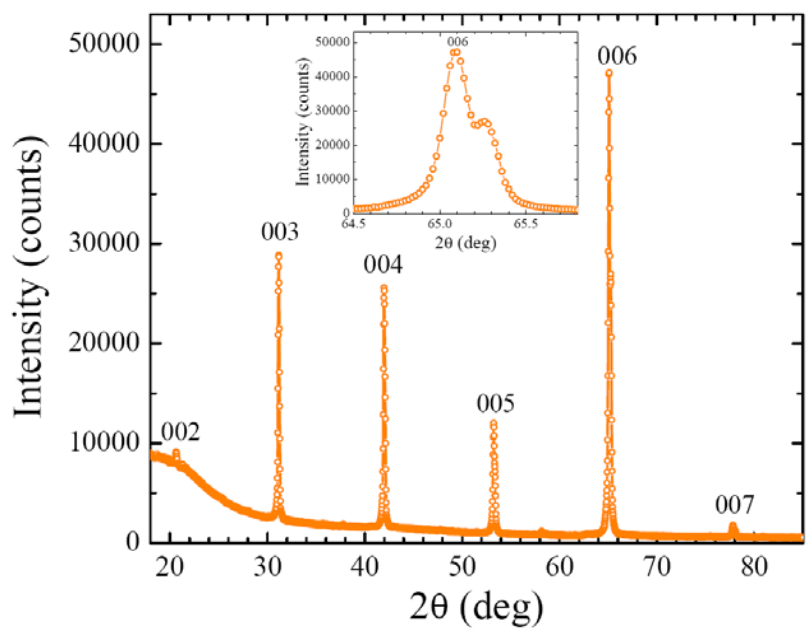

**Fig. 4.** X-ray diffraction pattern of several tens PrFeAs(O,F) whiskers showing the (00*l*) lines of the 1111-type structure. The inset shows an enlargement of the (006) diffraction peak.

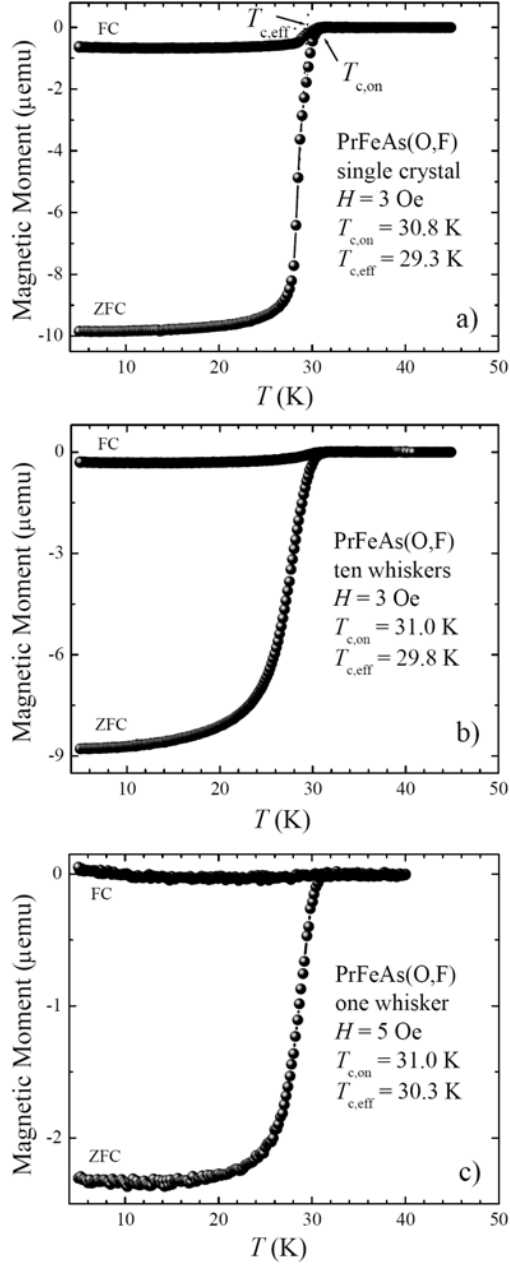

**Fig. 5.** Temperature dependences of the magnetic moment measured on one PrFeAs(O,F) single crystal with dimensions of ~ 120 × 700 × 1000 μm$^3$ (a), on ten single-crystal whiskers with dimensions in the range of ~ 5-6 × 50-60 × 700-1000 μm$^3$ collected from same single growth batch (b), and on one whisker with dimensions of ~ 7 × 70 × 600 μm$^3$ (c). The field was applied parallel to $c$-axis. ZFC and FC indicate zero-field cooling and field cooling curves, respectively. The transition temperature $T_{c,eff}$ and the transition onset temperature $T_{c,on}$ are marked by arrows.